# Absolute Properties of the Triple Star HP Aurigae

Claud H. Sandberg Lacy[1], Guillermo Torres[2], Marek Wolf[3], and Charles L. Burks[4]

[1]Physics Department, University of Arkansas, Fayetteville, AR 72701, USA;

clacy@uark.edu

[2]Harvard-Smithsonian Center for Astrophysics, 60 Garden Street, Cambridge, MA

02138; gtorres@cfa.harvard.edu

[3]Astronomical Institute, Faculty of Mathematics and Physics, Charles University in

Prague, 180 00 Praha 8, V Holesovickach 2, Czech Republic; wolf@cesnet.cz

[4]Physics Department, University of Arkansas, Fayetteville, AR 72701, USA;

clburks@email.uark.edu

ABSTRACT

New photometric, spectroscopic, and eclipse timing observations of the eclipsing binary star HP Aur allow for very accurate orbital determinations, even in the presence of a third body and transient starspot activity. The eclipsing binary masses are determined to an accuracy of ±0.4% and the radii to ±0.6%. The masses are 0.9543 ± 0.0041 and 0.8094 ± 0.0036 solar masses, and the radii are 1.0278 ± 0.0042 and 0.7758 ± 0.0034 solar radii, respectively. The orbital period in the outer orbit is accurately known for the first time: 4.332 ± 0.011 years. A comparison with current theories of stellar evolution show that the components' absolute properties can be well-matched by the current models at an age of about 7 billion years.

1. Introduction

Eclipsing binary stars have the potential to provide very accurate measurements of their components' masses, radii, luminosities and temperatures, among other fundamental stellar parameters, which may be used to critically test stellar evolution theory. In order to determine these parameters to very high accuracy, accurate and plentiful measurements are needed of the brightness variations over time (the light curve) and radial velocity variations over time (the radial velocity curve). Complications can arise that make more difficult the analysis of these fundamental data: the stars may have transient spots that cause the system's brightness to change in an irregular fashion, or there may be a faint third body in the system that causes the times of minima to vary in a cyclic manner over time. Indeed, in HP Aur we have both of these types of complications, but have nevertheless been able to accurately determine mean absolute properties for the components.

The detached main-sequence eclipsing binary star HP Aur (HD 280603, TYC 2401-1263-1) is a relatively bright star (V = 11.187 mag), originally classified as G0. It was first discovered as a variable star photographically by Strohmeier (1958). The first B and V-band photoelectric observations were made by Meinunger (1980), and several other light curves based on typically 200-300 photometric or CCD observations have been made since then. Meinunger measured the orbital period as 1.4228132 days. The first radial velocity measurements were those of Popper (2000). He summarized the state of the 5 existing photometric light curve studies at that time as "completely inadequate to evaluate the parameters of its light curve." Some of the photometric studies invoke orbital eccentricity (first asserted by Meinunger 1980), apsidal motion in an eccentric orbit (first asserted by Kozyreva et al. 1990), a faint third component with a 13.7 year period (Kozyreva et al. 2005), and spots/chromospheric activity (Liu et al. 1989, Eker et al. 2008).

Our current study is based on the analysis of 184 very accurate dates of minimum light, a large number (27) of new high-resolution spectrograms plus Popper's (2000) previously published ones (28), combined with a very large number (6685+2472) of new differential magnitudes obtained by two different and independent robotic telescopes. These new results are more definitive than in previous studies, and are accurate to better than 0.4% in the masses and 0.6% in the radii. The new and previously published dates of minimum light are discussed in section 2, the spectroscopic observations and reductions are discussed in section 3, the combined spectroscopic and eclipse timing analysis is discussed in section 4, the photometric study in section 5, and the absolute properties and comparison to theory in section 6.

2. Times of minimum light

Published times of minimum light as well as our own new dates are listed in Table 1. Only photoelectric or CCD results are given, since the older photographic data do not improve the fits significantly and just add noise. Our new previously unpublished dates of minima come from three observatories in the Czech Republic:  [NOTE FOR EDITOR: I was not able to figure out the Word equivalents for the LaTex that follows in this sentence.] Ondrejov (Ond\v{r}ejov Observatory, Czech Republic: 0.65-m reflecting telescope with an SBIG ST-8 or Apogee AP7p CCD camera and Johnson R filter), Ostrava (Observatory and Planetarium of Johann Palisa, Ostrava, Czech Republic: 0.20-m or 0.30-m telescopes with an SBIG ST-8XME CCD camera and Johnson R filter), and ValMez (Vala\v{s}sk\e Mezi\v{r}\'i\v{c}\'i Observatory, Czech Republic: 0.30-m Celestron Ultima telescope with an SBIG ST-7 CCD camera and Johnson VRI filters). Most of them were measured from images taken through the R filter. The weights originally assigned to the new dates were converted into mean uncertainties by calibrating the assigned weights of the new data with the published uncertainties of original observers of the other dates. These new dates and the other published ones are used in section 4 in a combined solution of dates of minima and radial velocities to estimate many orbital parameters and minimum masses.

3. Spectroscopic observations and reductions

HP Aur was observed spectroscopically at the Harvard-Smithsonian Center for Astrophysics (CfA) with an echelle instrument ("Digital Speedometer"; Latham 1992) attached to the 1.5m Tillinghast reflector at the Fred L. Whipple Observatory (Mount Hopkins, AZ). A total of 27 exposures were gathered between 2002 September and 2007 January with a resolving power of R ~ 35,000. A single echelle order was recorded with an intensified photon-counting Reticon detector giving a spectral range of 45 A centered at 5187 A, and including the lines of the Mg I b triplet. The signal-to-noise ratios of the exposures ranged from 16 to 30 per resolution element of 8.5 km/s. The wavelength solutions were based on exposures of a Th-Ar lamp before and after each science exposure.

Radial velocities were derived by using the two-dimensional cross-correlation algorithm TODCOR (Zucker & Mazeh 1994). The two templates (one for the more massive star and another for the less massive star) were selected from a library of synthetic spectra based on model atmospheres by R. L. Kurucz (see Nordstrom et al. 1994; Latham et al. 2002). The best match to the components was determined by seeking the maximum of the cross-correlation coefficient averaged over all exposures (see Torres et al. 2002). The template parameters that affect the velocities the most are the effective temperature and rotational broadening, while surface gravity and metallicity have a smaller effect. We adopted log g values of 4.5 for both components, which are close to our final results reported later, and we assumed solar metallicity. The optimal template match was found for temperatures of 5900 K and 5530 K for the more massive and less massive star, with estimated uncertainties of 100 K and 150 K, respectively, and rotational velocities of 41 ± 3 km/s and 30 ± 5 km/s.

The zero point of our velocity system was monitored regularly by means of exposures of the dusk and dawn sky, and small run-to-run corrections were applied as described by Latham (1992). Additionally, we performed numerical simulations as we have done in previous studies with similar spectroscopic material, to test for systematic effects caused by residual line blending as well as lines shifting in and out of our narrow spectral window as a function of orbital phase (see, e.g., Torres et al. 1997). These effects were found to be as large as 2 km/s for the more massive star and 6 km/s for less massive star, and corrections to the velocities were applied based on the simulations. These corrections increase the masses by about 6% for the more massive star and 4% for the less massive star. The final velocities in the heliocentric frame and individual uncertainties including all corrections are listed in Table 2. The median uncertainties are 1.5 km/s for the more massive star and 4.0 km/s for the less massive star.

The light ratio between the two stars was determined from our spectra following Zucker & Mazeh (1994). We obtained $L_2/L_1 = 0.31 +/- 0.02$ at the mean wavelength of the observations (5187 A), which is close to the V band.

The next section describes the evidence for a third body in the system, originally proposed by Kozyreva et al. (2005). We searched for lines of this star in our spectra using an extension of TODCOR to three dimensions (Zucker et al. 1995), but found no convincing evidence for it. We place an upper limit on its brightness of about 5% of the light of the more massive star, assuming it is a sharp-lined main-sequence star, or it could possibly be a white dwarf and would be undetectable by us.

An additional set of radial velocities of HP Aur was obtained by Popper (2000) between 1988 and 1997. They include corrections for systematics analogous to those we applied, and are of similar quality as ours, with formal uncertainties of 2.6 km/s for the more massive star and 2.5 km/s for the less massive star. These measurements were incorporated into our orbital solution described below in Sect. 4.

4.      Radial velocity/eclipse timing orbital solution

A diagram of the eclipse timing residuals from a linear ephemeris displays obvious periodic oscillations, particularly among the more recent timings that have better measurement errors. This strongly suggests a light travel time effect due to a third body in the HP Aur system. The radial velocities are likely affected at some level as well, which could potentially bias the mass determination. On the other hand, the velocities may contain useful information to help constrain the outer orbit, so we proceed here with a joint analysis of the eclipse timings and the velocities.

As a first step we carried out spectroscopic orbital solutions separately for the CfA and Popper velocities, to check for systematic differences. The agreement in the velocity amplitudes, which determine the masses, is excellent: the CfA observations give $K_1 = 104.61 \pm 0.35$ km/s and $K_2 = 124.08 \pm 0.93$ km/s, while those of Popper yield $K_1 = 104.42 \pm 0.59$ km/s and $K_2 = 124.14 \pm 0.56$ km/s. The resulting minimum masses differ by less than 0.3%. We did find an apparently significant difference in the center-of-mass velocities (which are $+18.18 \pm 0.29$ km/s for CfA and $+19.94 \pm 0.38$ km/s for Popper) that could be instrumental in nature, but this could also be due to orbital motion in the outer orbit.

We then combined both velocity data sets with the eclipse timings into a global solution, solving simultaneously for the elements of the inner and outer orbits (twelve free parameters). We included one additional free parameter to represent a possible difference between the instrumental zero points of the two velocity data sets, for a total of thirteen unknowns. The formulation for the effect that the third body has on the timings follows the work of Irwin (1952, 1959), and the joint solution was carried out using standard non-linear least-squares techniques (e.g., Press et al. 1992). The dates of the velocity observations were corrected for light travel time during the iterations. Individual weights were applied to all observations based on the internal errors. The eclipse timings with no published uncertainties were assigned an error (separately for the primary eclipse and secondary eclipse measurements) such that the chi-square per degree of freedom for those measurements was close to unity. The formal uncertainties of the other data sets were adjusted by a multiplicative scale factor to achieve a similar condition. This was done separately for the more massive star and less massive star velocities from CfA and Popper (2000), as well as for the primary and secondary eclipse timings with published errors. The adjustment factors for the errors of the eclipse timings were 3.1 for the primary eclipses and 2.7 for the secondary eclipses. Those for the uncertainties of the CfA velocities were close to unity. The original errors of the Popper velocities, on the other hand, were found to be *overestimated* by factors of 2.0 and 2.3 for the more massive star and less massive star velocities, respectively. This is explained by the fact that those errors were based on the rms residuals from Popper's orbital fits, which did not account for perturbations caused by the third star, not known at the time. This resulted in an increased scatter (see below), noted also by Popper.

The elements we obtained for the inner and outer orbits are presented in Table 3. The outer orbit is fairly eccentric and has a period of about 4.3 yr, which is *very different* from the value of 13.7 yr proposed by Kozyreva et al. (2005). For our final fit we considered the inner orbit to be circular. Test solutions in which we allowed the eccentricity to vary gave a value not significantly different from zero. Similarly, the offset between the zero points of the two velocity data sets was negligible when allowed to vary, so in the final fit we set it to zero. The figures below display the results.

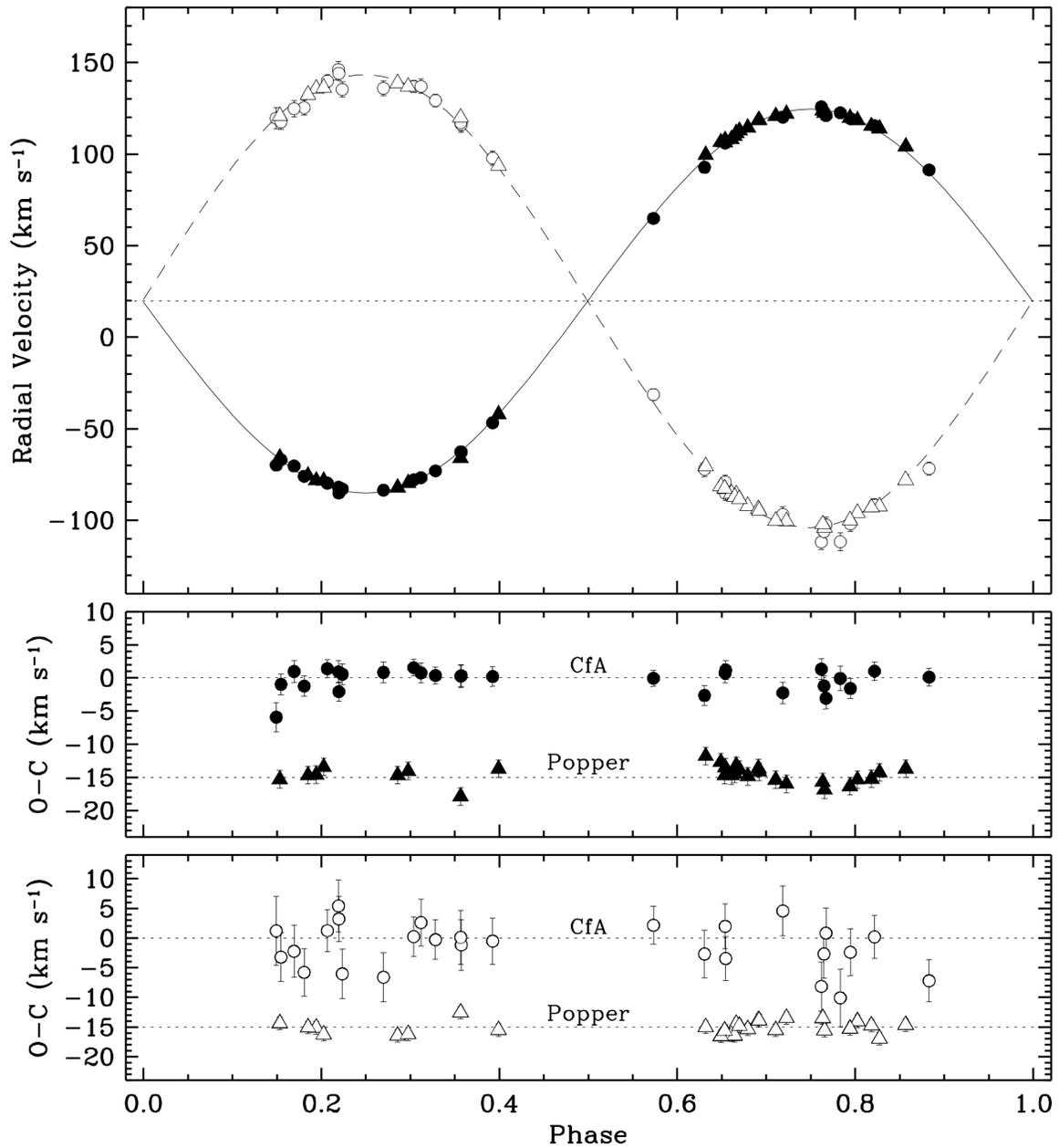

Figure 1: Radial-velocity measurements for HP Aur from CfA and from Popper (2000) shown with circles and triangles, respectively, along with our best fit model. Filled

symbols and the solid line correspond to the more massive star 1. The dotted line represents the center-of-mass velocity of the triple system. Motion in the outer orbit has been subtracted from the measurements and from the model in the top panel. Residuals are shown in the bottom panels, with those from Popper displaced vertically for clarity. The more massive star 1 residuals from Popper (2000) show signs of a systematic trend, which is however not present in those of the less massive star 2.

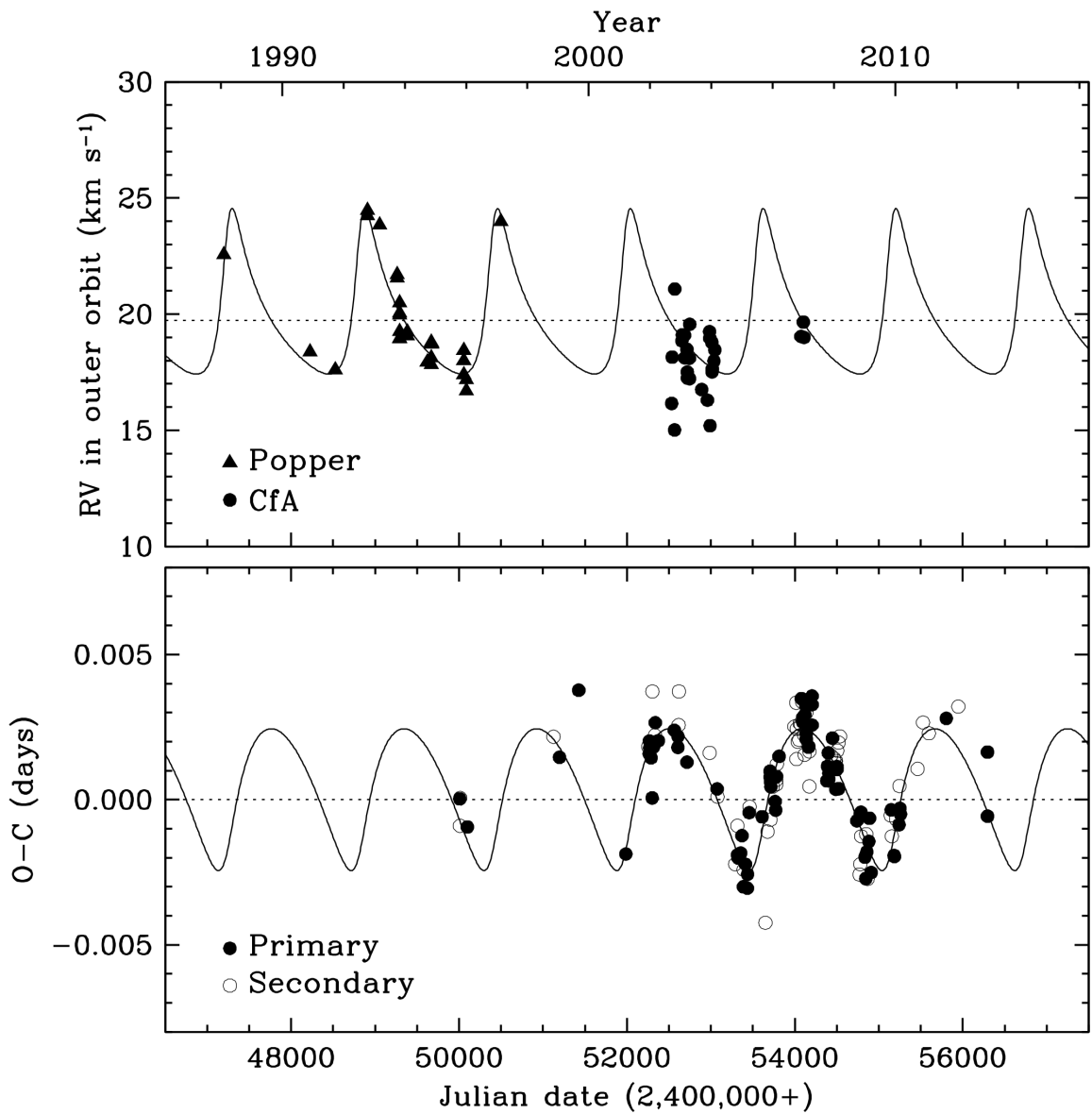

Figure 2: Motion of the center of mass of the eclipsing pair in the HP Aur system as traced by the radial velocity and eclipse timing observations obtained in the last 15 years, along with our model for the outer orbit. A few older eclipse timings are not shown to avoid compressing the time scale. Top: The symbols correspond to the mean radial velocity of the inner binary's center of mass, computed as a weighted average of the primary and secondary velocities after subtracting the motion in the 1.4 day orbit. The larger dispersion for CfA is driven by the larger residuals of the secondary velocities in that data set compared to the Popper observations (see Table 3), when taking the average. The dotted line represents the center-of-mass velocity of the triple. Bottom: The bottom panel shows the light travel time curve with the O-C residuals from the linear ephemeris given in Table 2. The semi-amplitude of this effect corresponds to about 3.7 minutes.

As may be seen in Figure 2, the CfA velocities were all obtained over a relatively narrow range of phases in the outer orbit. Therefore they are not much affected by the motion in the outer orbit, but also do not constrain it very much. The observations by Popper (2000), on the other hand, span two full cycles of the outer orbit and sample most phases well. This explains why they are more affected by the changing center of mass of the eclipsing pair, which led Popper to overestimate his errors that were based on the dispersion from a double-lined orbital fit. Conversely, these velocities provide strong constraints on the outer orbit that complement the timing measurements very well, as they cover an interval of time in which few eclipses were observed.

5.      Photometric study

Differential photometry was produced from images taken by two independent robotic telescopes. The URSA WebScope is built from a Meade 10-inch f/6.3 LX-200 telescope with a Santa Barbara Instruments Group ST8 CCD camera (binned 2x2 to produce 765x510 pixel images having 2.3 arcsec square pixels) inside a Technical Innovations Robo-Dome, and controlled with an Apple Macintosh G4 computer executing a nightly observing schedule. The observatory is located atop Kimpel Hall on the Fayetteville, Arkansas campus, with the control room located beneath the observatory inside the building. 120 second exposures through a Bessel V filter (2.0 mm of GG 495 and 3.0 mm of BG 39) were read out and downloaded to the control computer over a 30 second interval, then the next exposure was begun. The observing cadence was about 150 s per observation. The variable star would often be monitored continuously for 1- 6 hours. HP Aur was observed by URSA on 132 nights during parts of 10 observing seasons from 22 Nov. 2001 to 23 Sept. 2010, yielding 6685 observations.

The other telescope we used is the NFO WebScope, a refurbished 24-inch Group 128 cassegrain reflector with a 2K x 2K Kodak CCD camera, located near Silver City, NM (Grauer, Neely, & Lacy 2008). Observations consisted of 120 second exposures through a Bessel V filter. HP Aur was observed by NFO on 68 nights during parts of 4 observing seasons from 29 Jan. 2005 to 27 Feb. 2008, yielding 2472 observations.

The images were flat-fielded and measured by an automated application (Measure, written by author Lacy). Extinction coefficients were determined nightly from the comparison star measurements. They averaged 0.26 mag/airmass at URSA (they ranged from 0.19 to 0.31 mag/airmass), 0.19 mag/airmass at the NFO (they ranged from 0.13 to 0.26 mag/airmass). The comparison stars were TYC 2401-0760-1 (BD +69°

1005, $V_T = 10.695$ mag) and TYC 2401-0224-1 (HD 280491, $V_T = 10.343$ mag, K2).

Both comparison stars are west of the variable star, but within 8 arcmin of it. The mean

nightly magnitude difference between the comparison stars was constant within a

standard deviation of 0.028 mag (URSA) and 0.013 mag (NFO). The scatter in

comparison star magnitude differences within a night averaged 0.007 mag (URSA) and

0.005 mag (NFO) for the standard deviation of the nightly differential magnitudes. For

the differential magnitudes, URSA data uses the variable minus the first comparison star

only, whereas for NFO data, the sum of the fluxes of both comparison stars was

converted to a magnitude called "comparisons". The resulting 6685 (URSA) and 2472

(NFO) V magnitude differences (variable-comparison) or (variable-comparisons) are listed in Tables 4 & 5 (without any nightly magnitude corrections or light travel-time corrections) and are shown in Figures 4-6 (after the nightly magnitude corrections and light travel-time corrections discussed below have been added). In terms of time coverage, the NFO observations lie in the middle half of the URSA observational coverage.

5.1    Photometric orbit

Preliminary examination of the light curves showed relatively large (several

hundredths of a mag) night-to-night and longer-period variations in the URSA and NFO

data, ranging from nights to years, whereas this had never occurred before with URSA

light curves of many other eclipsing binary stars, so it is clearly intrinsic to the variable star itself. The light curve fitting was done with the Nelson-Davis-Etzel (NDE) model as implemented in the code *jktebop* (Etzel 1981; Popper & Etzel 1981, Southworth et al.

2007). An initial fit was made in order to get a preliminary mean light curve, from which it was possible to determine the mean residuals for each night. These mean residuals (observed – computed) are plotted in Figure 3. The binary appears to have become about 0.1 mag fainter over the last decade. The observed variations due to transient starspots are so rapid, random, and numerous that we cannot hope to model them accurately with our data set, but by subtracting off the mean of the nightly residuals from the preliminary light curve, we can get a mean light curve that will yield reliable mean parameters for further analysis.

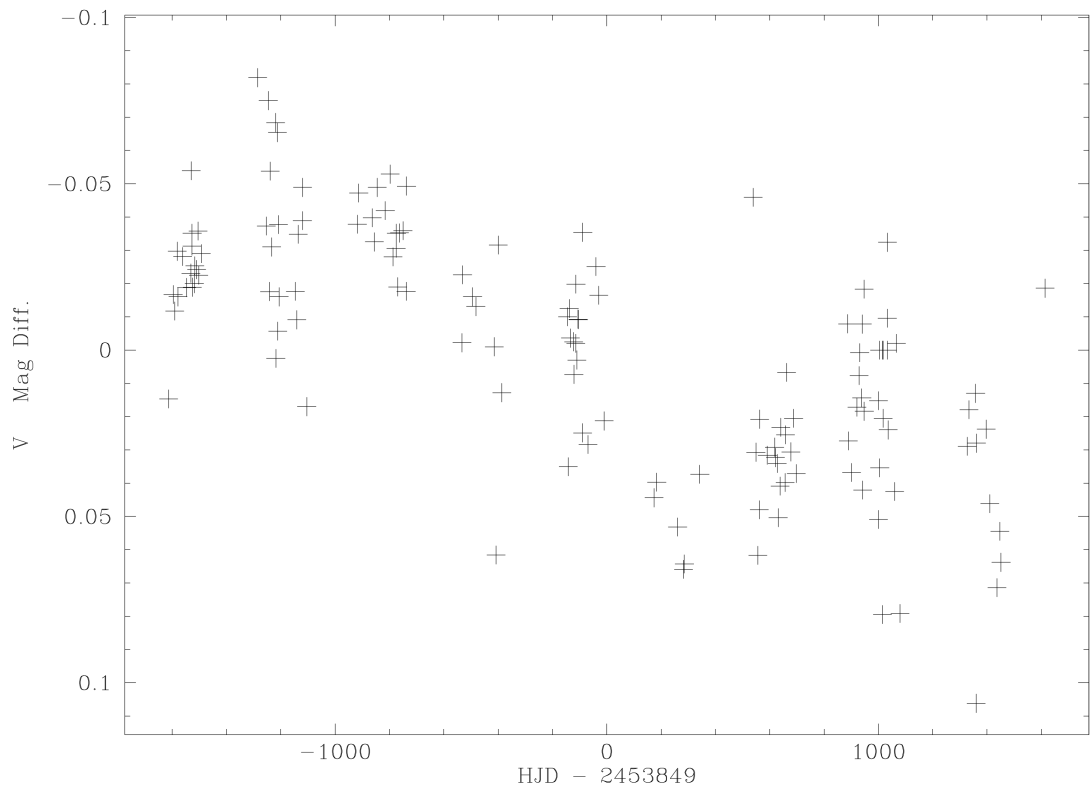

Figure 3: Variations in the light level of HP Aur over a 9-year interval. Variations are seen over periods of days to years. The star has dimmed by roughly 0.1 mag over about a decade.

The light curves that have the nightly residuals removed and also the light travel-time lags around the 3-body center of mass removed are shown in Figures 4-6.

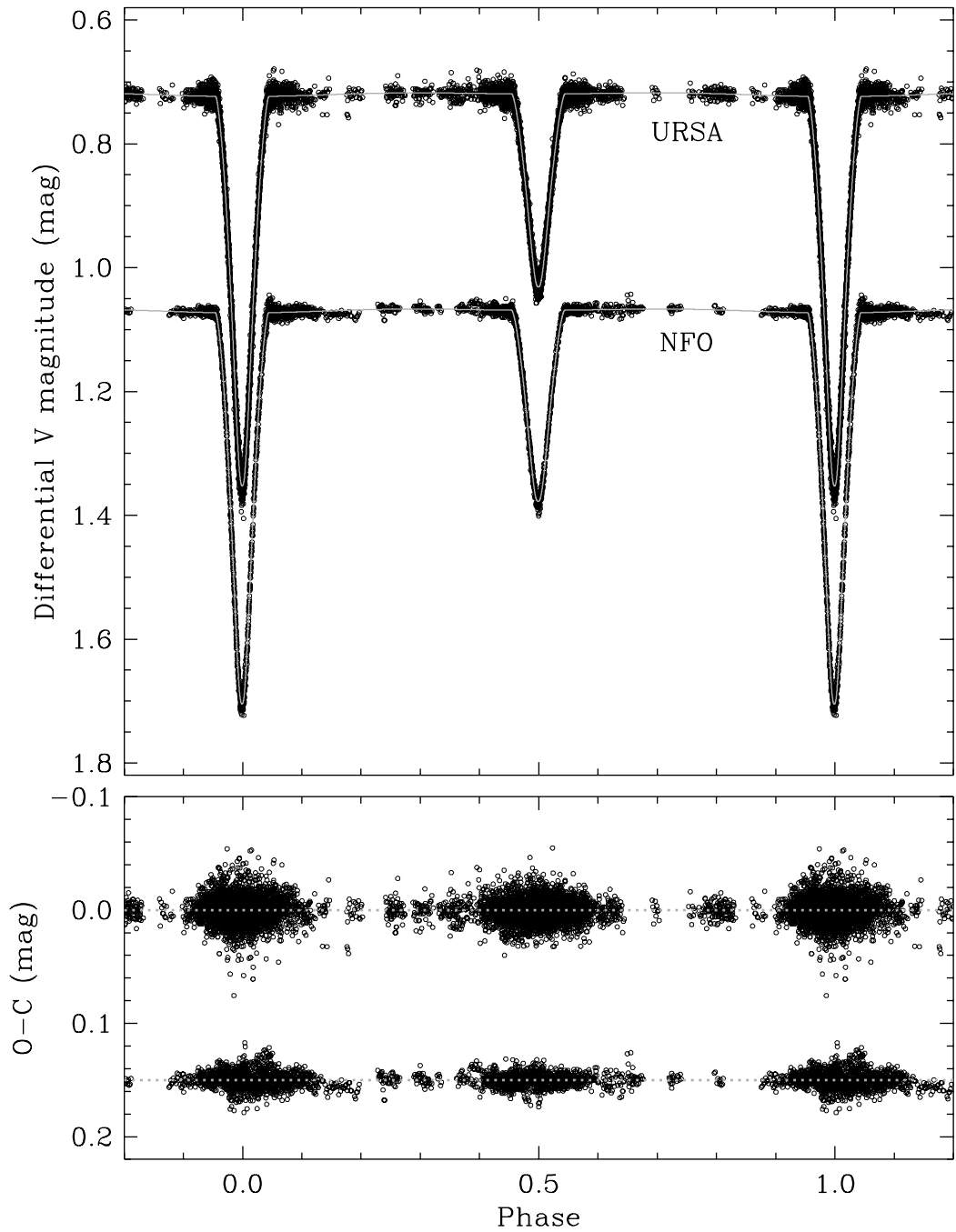

Figure 4: URSA and NFO mean light curves after removal of mean nightly residuals and time lags due to 3-body motion around the center of mass.

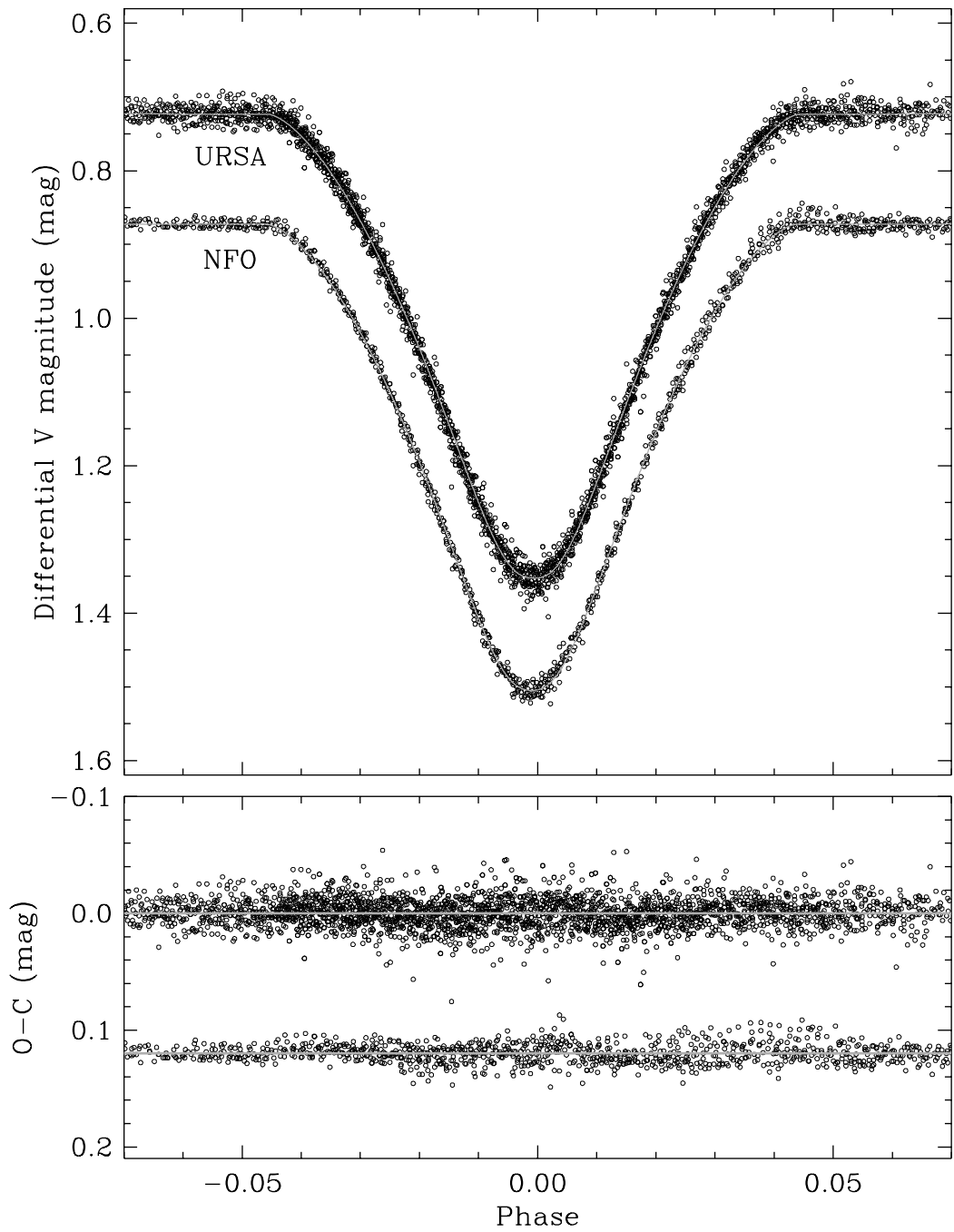

Figure 5: Primary (deeper) eclipse of HP Aur.

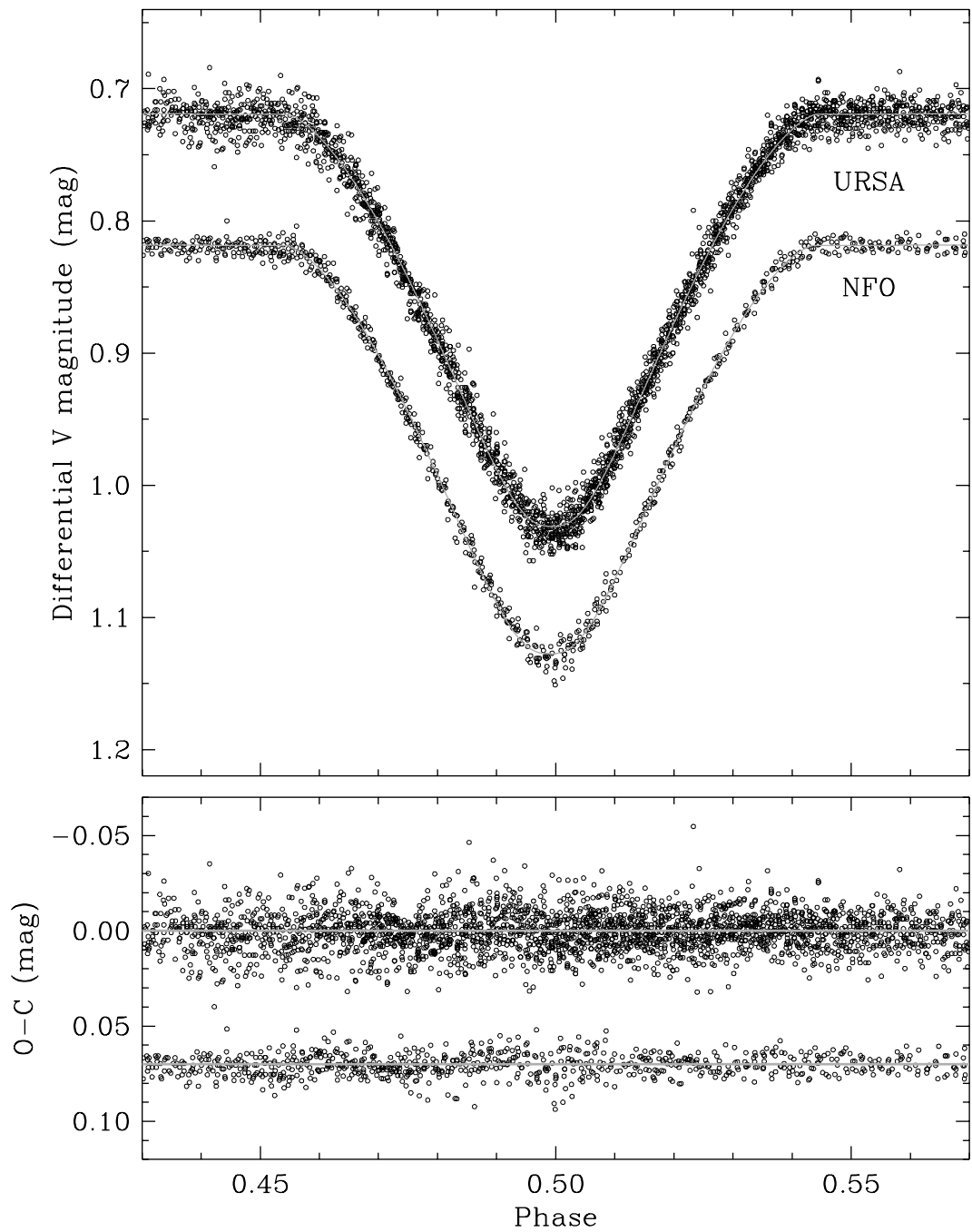

Figure 6: Secondary (shallower) eclipse of HP Aur.

Photometric orbital elements resulting from the fits are given in Table 6. The principal fitted elements are $J_2(V)$, the visual central surface brightness of the cooler, smaller, less massive star 2, $r_1+r_2$, the sum of relative radii of the stars, k, the ratio of the smaller to larger radius, i, the orbital inclination in degrees, $u_1$ and $u_2$, the limb-darkening coefficients. Auxilliary quantities related to the fit include $L_1$ and $L_2$, the normalized luminosities of the stars, $\sigma$, the standard error of an observation, N, the number of observations, and the number of Corrections applied to the instrumental data. The value of the mass ratio was fixed at the spectroscopically fitted value. In finding the photometric solutions we initially assumed that the limb-darkening coefficients would be near values predicted by atmospheric theory for stars of approximately the temperatures of ours, about 0.67 and 0.74 for the coefficients, and we fixed their values in the solutions. These solutions, however, had light ratios, $L_2/L_1$, near 0.6, which was far from the spectroscopic value, $0.31 \pm 0.02$ at V. Eventually it was decided that, because these stars were known from our observations of the variations in brightness over time to have spots and were reported as chromospherically active, the limb-darkening coefficients were likely to be non-standard, so we let them be variables in the solutions. The solutions then converged to values with light-ratios near $0.318 \pm 0.004$ for limb-darkening values near 0.15 and 0.09. These unusually small limb-darkening values underscore the facts that this system has transient spots and chromospheric activity. The two independent sets (URSA & NFO) of fitted photometric orbital parameters do agree with each other remarkably well considering that they were derived from different telescopes at different times, and considering there are spot variations on a number of time scales in the raw data.

We have experimented with fitting orbits to some other much smaller photometric data sets for HP Aur, those of Meinunger (1980, B & V), Kozyreva (1990, V), and Wolf (unpublished, R).  The general result is that the smaller data sets will not converge with the limb-darkening coefficients left as free variables because there is not enough high-quality data in the sets.  By fixing the limb-darkening parameters to values near those derived from the URSA & NFO sets, and then creating grids of solutions by varying the value of the ratio of radii, k, we can find solutions that have luminosity ratios $L_2/L_1$ in the range allowed by the independently measured spectroscopic light ratio.  There are a few parameter differences that may possibly be significant, but the general results of the light curve analyses are similar between the various smaller data sets and the URSA and NFO data sets.  Because the URSA and NFO sets are so much larger, their parameter uncertainties are very much smaller than the other sets, and we have chosen to use only these best data sets in our final analyses.

Lacy (1987) showed that the difference in visual surface brightness parameter, $\Delta F_v$, is connected to the normalized V-band central surface brightness of the cooler star in eclipsing binaries ($J_1 = 1$): $\Delta F_v = 0.25 \log J_2$'.  Here, $J_2$ is a fitted parameter in the *jktebop* code that models the light curves, and $J_2$' is that value corrected for differences in limb-darkening parameters.  Popper's (1980) Table 1 gives the relationship between the visual surface brightness parameter $F_v$ and the stellar temperature T, thus the difference in temperature $\Delta T$ is readily and very accurately determined from the V-filter light curve fit alone.  Additional light curves in different bandpasses are not necessary to determine accurately the temperature difference determined by this surface-brightness method.

Tests for third light in both data sets showed that it is not detectable at a significant level. The radial velocity and eclipse timing data showed that the eccentricity in the short-period orbit is zero within the uncertainties, so that parameter was also fixed at zero in the final analyses.

The NDE model used by *jktebop* has been compared with more complicated models (Popper & Etzel 1981; North & Zahn 2004) including the WD model.  The principal results of these studies are that the limits for high-accuracy measurements of parameters such as the radii, inclination, etc. with the NDE model are that the component oblateness should be less than 0.04 and the mean radii should be less than 0.25.  Since the HP Aur properties are all well within these limits, we do not feel the need to use a more complicated model in this case.

6.  Absolute properties of HP Aur

Our measured value of $J_2$ from the light curves enables a more precise determination of the temperature difference than that allowed by the spectroscopy (see the discussion about $\Delta F_v$ above).  We obtain $660 \pm 30$ K by using the visual flux calibration by Popper (1980). Our spectroscopic temperature for the more massive star of 5900 K then implies 5240 K for the less massive star. We note, however, that our spectroscopic estimates depend rather strongly on the adopted metallicity, for which we used the solar value in Sect. 3 as no other spectroscopic determination is available. An additional measure of the temperature that is much less sensitive to metallicity may be obtained from standard photometry available in the literature for the combined light, along with color-temperature calibrations from Casagrande et al. (2010). We used the following sources for the photometry: $B_T$ and $V_T$ from the Tycho-2 Catalog (Hog et al. 2000), B and V from the APASS catalog (Henden et al. 2012), $JHK_s$ from the 2MASS catalog (Cutri et al. 2003), V and $I_C$ from the TASS catalog (Droege et al. 2006), B and V from Kozyreva et al. (2005), and a B-V index from Liu et al. (1989). We constructed 10 different but non-independent color indices, and de-reddened them following Cardelli et al. (1989). An estimate of the reddening was obtained by consulting the dust maps by Schlegel et al. (1998), Drimmel et al. (2003), and Amores & Lepine (2005), from which we derived E(B-V) values of 0.042, 0.039, and 0.064 mag, respectively, for an assumed distance of 210 pc (see below). The average reddening adopted is E(B-V) = 0.048 ± 0.020, with a conservative uncertainty. We obtained a mean temperature for HP Aur of 5560 ± 100 K. Assuming this corresponds to the luminosity-weighted average of the two components, the flux scale of Popper (1980) results in a temperature of 5730 K for the more massive star and a value of 5100 K for the less massive star. Finally, we averaged our two estimates for the more massive star to obtain 5810 ± 120 K, and derived the temperature of the less massive star as 5160 ± 120 K from a slightly adjusted temperature difference based on $J_2$ (650 ± 30 K). The corresponding spectral types are approximately G2 and K1. The final temperature uncertainties for the more massive and less massive stars include a contribution of 100 K added in quadrature to account for possible systematics in our spectroscopic and photometric estimates. The absolute properties of HP Aur are given in Table 7.

A distance to HP Aur of 210 ± 15 pc was obtained by using the visual surface brightness parameter calibration of Popper (1980) and the average out-of-eclipse brightness (V = 11.187 ± 0.018) corrected for extinction. This distance estimate is in good agreement with an independent one involving bolometric corrections from Flower (1996) (see also Torres 2010).

The measured projected rotational velocities appear to be roughly consistent with the predicted synchronous values listed in Table 7. The $M_3 \sin i$ value in Table 3 along with the absolute masses of the more massive and less massive eclipsing stars lead to an estimated minimum mass for the tertiary of about 0.27 $M_{Sun}$, corresponding to a late M dwarf (SpT approximately M4). On the other hand, from our spectroscopic limit on the brightness of this star in Sect. 3, its spectral type is not likely to be earlier than about M0 if it is a main-sequence star, implying an upper limit on the mass of around 0.5 $M_{Sun}$ (unless it is a white dwarf).

7.     Comparisons with theory

The accurate masses and radii of the eclipsing binary in HP Aur (see Table 7), along with the temperatures, permit a meaningful comparison with stellar evolution theory. To be conservative, the error estimates on the radii in Table 6 have been doubled since the stars are variable over time.  Figure 7 shows the observations against models from the Yonsei-Yale series (Yi et al. 2001).

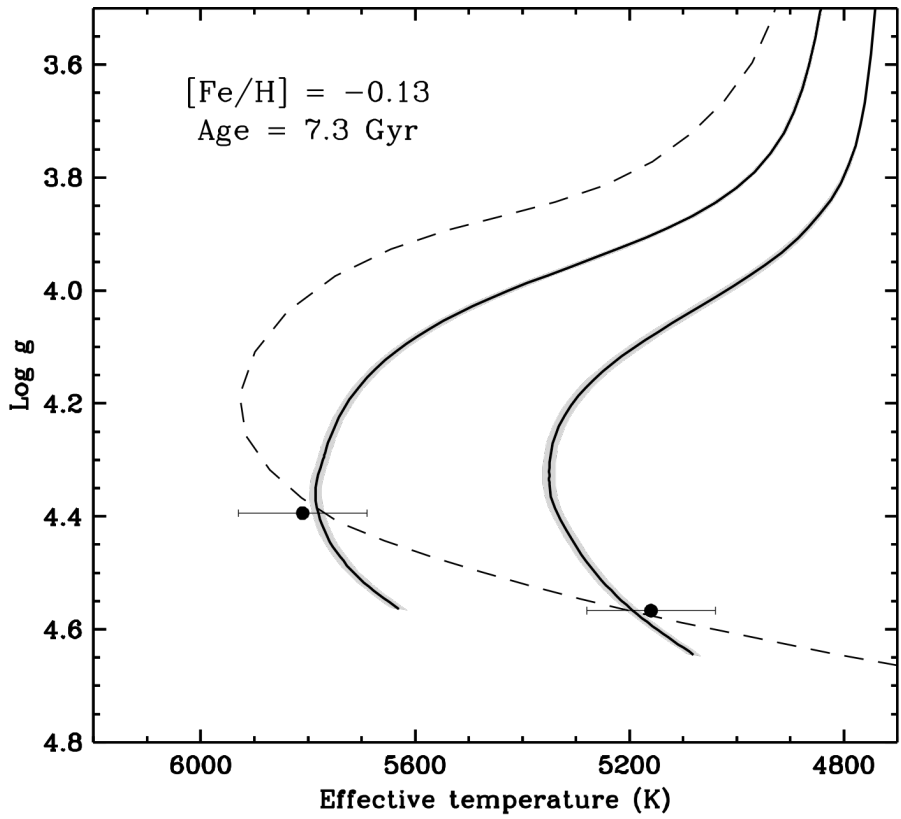

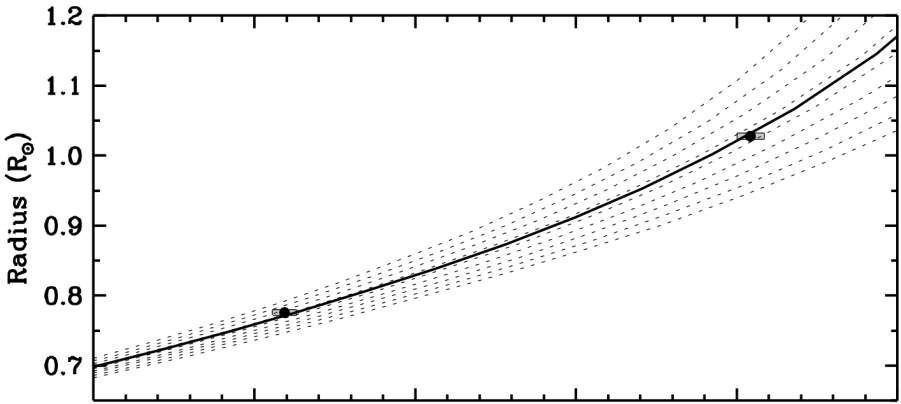

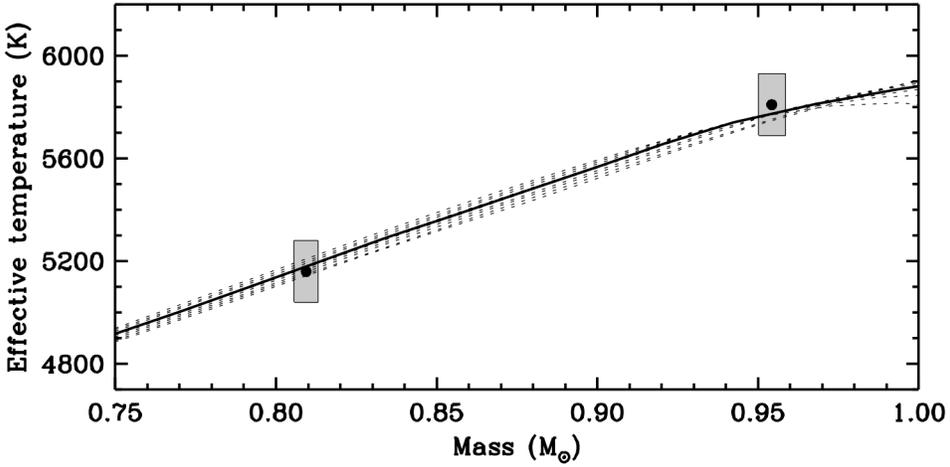

Figure 7: Observations of HP Aur compared against Yonsei-Yale models (Yi et al. 2001). Top: The solid lines show evolutionary tracks for the measured masses, with the gray areas around each track representing the uncertainty in their location that comes from the mass errors. The best-fit isochrone for a metallicity of [Fe/H] = -0.13 (Z = 0.0136) and an age of 7.3 Gyr is shown with a dashed line. Middle and bottom: Radius and temperature as a function of mass. Isochrones for the same metallicity as above and ages between 5 and 9 Gyr are shown with dotted lines (in 0.5 Gyr steps), and the best-fit 7.3 Gyr isochrone is indicated with a thick solid line.

The top panel displays evolutionary tracks for the measured masses, where we have adjusted the metal abundance of the models to match our estimated effective temperatures. The best-fit metallicity is Z = 0.0136, which in these models corresponds to [Fe/H] = -0.13 (assuming [α/Fe] = 0.0). All properties of both components are well fit at an age of 7.3 Gyr. The corresponding isochrone is shown with a dashed line. The lower panels of Figure 7 illustrate the agreement in radius and in temperature, with the solid line representing the same isochrone as in the top panel. The more massive star of the eclipsing binary has evolved considerably, and in terms of its age it is about 3/4 of the way to exhausting its central hydrogen. According to these models the central helium abundance is $Y_c = 0.78$ (from a starting fraction of about $Y_c = 0.257$ at this metallicity). The lower-mass star, on the other hand, is only about 1/3 of the way to the terminal-age main sequence, and currently has a central helium fraction of $Y_c = 0.54$.

The Yonsei-Yale models adopt gray model atmospheres as boundary conditions between the photosphere and the interior. For stars of solar type or hotter this is a good approximation, but for cooler stars it has been shown that non-gray model atmospheres are needed for a more realistic representation of stellar structure (see, e.g., Chabrier & Baraffe 2000, and references therein). The less massive K1 star in HP Aur is cool enough that this may affect the comparison described above. We have therefore made an additional comparison using models from the Dartmouth series (Dotter et al. 2008), which use non-gray boundary conditions as well as a more sophisticated equation of state. Figure 8 shows that the Dartmouth models also provide an excellent fit to the observations, for a metallicity of [Fe/H] = -0.06 and an age of 7.0 Gyr which are similar to those found earlier. All measured quantities agree with predictions to well within their uncertainties, including the well measured temperature difference.

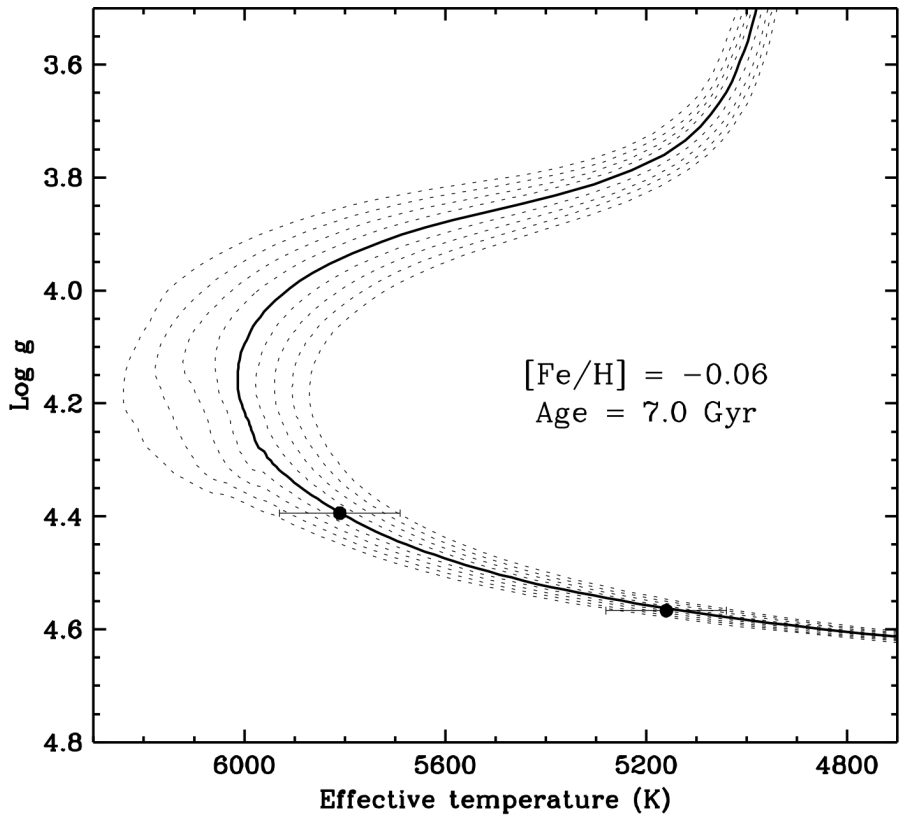

[Fe/H] = −0.06
Age = 7.0 Gyr

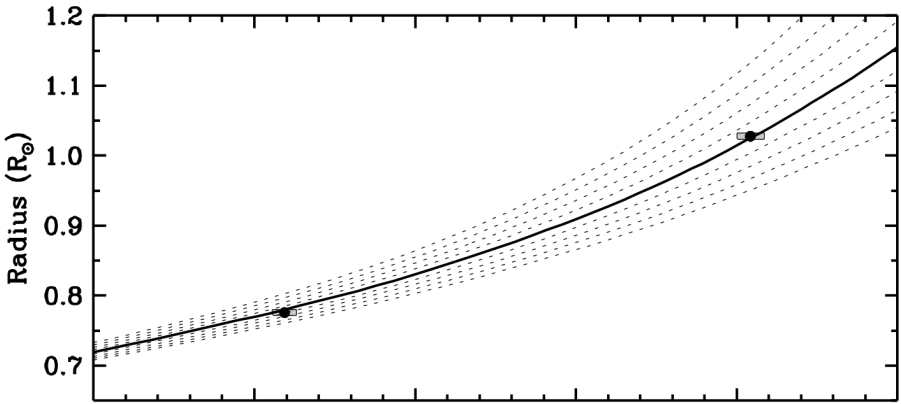

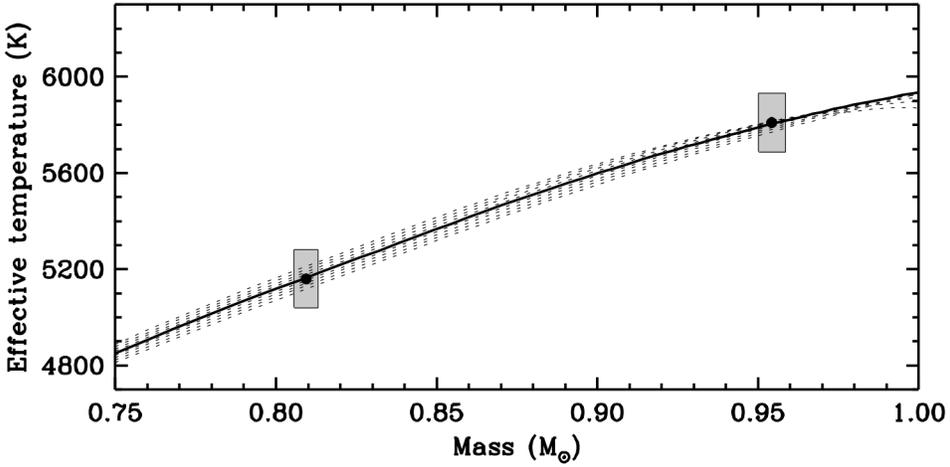

Figure 8: Similar to Figure 7 for the Dartmouth models of Dotter et al. (2008). Top: Isochrones in 0.5 Gyr steps from 5 to 9 Gyr (left to right) are shown with dotted lines. The best-fit 7.0 Gyr model is represented with a solid line. Middle and bottom: Same isochrones as above in the radius versus mass and temperature versus mass diagrams.

The less massive star of HP Aur is near the regime where other low-mass stars in short-period binaries have shown discrepancies with stellar evolution models, in the sense of being larger and cooler than predicted. Objects displaying these disagreements have generally been found to be rapid rotators and to have high levels of chromospheric activity. HP Aur was detected as a strong X-ray source by ROSAT (Voges et al. 1999). Based on the measured count rate and hardness ratio HR1 we estimate its total X-ray luminosity to be $Log(L_X) = 30.4$, and the X-ray to bolometric luminosity ratio to be $log(L_X/L_{bol}) = -3.35$. This value is near the observed saturation level for coronal X-ray sources of $log(L_X/L_{bol}) \sim -3$ (see, e.g., Pizzolato et al. 2003), indicating the system is indeed very active, as might be expected from its short orbital period. Assuming the stars in HP Aur are rotationally synchronized, as appears to be the case, the Rossby numbers of the components (defined as the ratio between the rotation period and the convective turnover time) inferred from their spectral types ($Log R_0 = -1.2$ and $-1.4$ for the more massive and less massive stars) place both components in the range where stars usually present photometric variability due to spots (see Hall 1994). This is consistent with what we see in the light curves of HP Aur. Furthermore, both stars show the Ca II H and K lines in emission, and H$\alpha$ is in absorption but filled in (Strassmeier et al. 1993).

Our model comparisons above give no indication of any significant discrepancies in radius or in temperature for the lower-mass star of the kind exhibited by other active stars, which is somewhat unexpected. Also unusual is the fact that the system appears quite old (~7 Gyr) despite being very active. The indications that the more massive star is also an active star suggest the possibility that *both* components may be inflated and may have their temperatures suppressed by the high activity levels, but with anomalies that are similar in magnitude so that the stars still fall on a single isochrone. The larger mass of the more massive star (only 5% smaller than the Sun's) does not necessarily exempt it from possibly having these effects, which have been reported in a number of similarly active and well measured stars of about the same mass (e.g., V1061 Cyg B, FL Lyr B, CV Boo B, ZZ UMa B, EF Aqr B, all with M > 0.93 M_Sun; Torres et al. 2006, Torres et al. 2008, Clausen et al. 2009, Vos et al. 2012). We are unable to explore this hypothesis further with the evidence in hand because our model comparisons have two free parameters (age and metallicity), the combination of which can give a wide range of slopes for the mass-radius and mass-temperature relations. The issue could be resolved, however, with an accurate spectroscopic determination of the metallicity of HP Aur to remove the degeneracy.

The authors wish to thank Bill Neely who operates and maintains the NFO WebScope for the Consortium, and who handles preliminary processing of the images and their distribution. We thank P. Berlind, M. Calkins, G. Esquerdo, and D. W. Latham for help in obtaining the CfA spectra, and R. J. Davis for maintaining the CfA echelle database over the years. GT acknowledges partial support for this work from NSF grant AST-1007992. The research of MW was supported by the Research Program MSM0021620860 "Physical Study of objects and processes in the Solar System and in

Astrophysics" of the Ministry of Education of the Czech Republic. [NOTE TO EDITORS: the observers' names that follow are in LaTeX because they are in Czech.] We also thank the amateur observers from Ostrava and Valasske Mezirici: K. Ho\v{n}kov\'a, J. Jur\'y\v{s}ek, and L. \v{S}melcer.

Figure Captions:

Figure 1: Radial-velocity measurements for HP Aur from CfA and from Popper (2000) shown with circles and triangles, respectively, along with our best fit model. Filled symbols and the solid line correspond to the more massive star 1. The dotted line represents the center-of-mass velocity of the triple system. Motion in the outer orbit has been subtracted from the measurements and from the model in the top panel. Residuals are shown in the bottom panels, with those from Popper displaced vertically for clarity. The more massive star 1 residuals from Popper (2000) show signs of a systematic trend, which is however not present in those of the less massive star 2.

Figure 2: Motion of the center of mass of the eclipsing pair in the HP Aur system as traced by the radial velocity and eclipse timing observations obtained in the last 15 years, along with our model for the outer orbit. A few older eclipse timings are not shown to avoid compressing the time scale. Top: The symbols correspond to the mean radial velocity of the inner binary's center of mass, computed as an average of the more massive star 1 and less massive star 2 velocities after subtracting the motion in the 1.4 day orbit. The dotted line represents the center-of-mass velocity of the triple. Bottom: The bottom panel shows the light travel time curve with the O-C residuals from the linear ephemeris given in Table 2. The semi-amplitude of this effect corresponds to about 3.7 minutes.

Figure 3: Variations in the light level of HP Aur over a 9-year interval. Variations are seen over periods of days to years. The star has dimmed by roughly 0.1 mag over about a decade.

Figure 4: URSA and NFO mean light curves after removal of mean nightly residuals and time lags due to 3-body motion around the center of mass.

Figure 5: Primary eclipse of HP Aur.

Figure 6: Secondary eclipse of HP Aur.

Figure 7: Observations of HP Aur compared against Yonsei-Yale models (Yi et al. 2001). Top: The solid lines show evolutionary tracks for the measured masses, with the gray areas around each track representing the uncertainty in their location that comes from the mass errors. The best-fit isochrone for a metallicity of [Fe/H] = -0.13 (Z = 0.0136) and an age of 7.3 Gyr is shown with a dashed line. Middle and bottom: Radius and temperature as a function of mass. Isochrones for the same metallicity as above and ages between 5 and 9 Gyr are shown with dotted lines (in 0.5 Gyr steps), and the best-fit 7.3 Gyr isochrone is indicated with a thick solid line.

Figure 8: Similar to Figure 7 for the Dartmouth models of Dotter et al. (2008). Top: Isochrones in 0.5 Gyr steps from 5 to 9 Gyr (left to right) are shown with dotted lines.

The best-fit 7.0 Gyr model is represented with a solid line. Middle and bottom: Same isochrones as above in the radius versus mass and temperature versus mass diagrams.

Table 1. Dates of minimum light for HP Aur

| Type of Eclipse[a] | HJD-2400000 | Precision (days) | Ref. |
|---|---|---|---|
| | | | |
| 1 | 42756.3475 | 0.0020 | 1 |
| 2 | 42805.4375 | 0.0020 | 1 |
| 2 | 43498.3465 | 0.0020 | 1 |
| 1 | 43577.3113 | 0.0020 | 1 |
| 2 | 43848.3595 | 0.0020 | 1 |
| 1 | 43850.4903 | 0.0020 | 1 |
| 1 | 46008.9135 | 0.0020 | 2 |
| 2 | 46013.8954 | 0.0020 | 2 |
| 1 | 46018.8732 | 0.0020 | 2 |
| 1 | 46041.6383 | 0.0020 | 2 |
| 1 | 46045.9068 | 0.0020 | 2 |
| 1 | 46068.6719 | 0.0020 | 2 |
| 2 | 46070.8100 | 0.0020 | 2 |
| 1 | 46353.2355 | 0.0004 | 3 |
| 2 | 46355.3694 | 0.0003 | 3 |
| 1 | 50008.4580 | 0.0002 | 3 |
| 2 | 50010.5913 | 0.0001 | 3 |
| 2 | 50013.4379 | 0.0004 | 3 |
| 1 | 50099.5175 | 0.0001 | 3 |
| 2 | 51080.5584* | 0.0001 | 4 |
| 2 | 51124.6623 | 0.0003 | 4 |
| 1 | 51196.5140 | 0.0001 | 4 |
| 1 | 51425.5903 | 0.0010 | 5 |
| 1 | 51876.6186 | 0.0003 | 6 |
| 2 | 51901.517 | 0.0010 | 6 |
| 2 | 51935.6673 * | 0.0001 | 7 |
| 1 | 51937.7984 * | 0.0001 | 8 |
| 1 | 51983.3303 | 0.0003 | 6 |
| 1 | 51987.5985 | 0.0002 | 8 |
| 2 | 52005.385 | 0.0005 | 6 |
| 2 | 52217.38704 | 0.0005 | 6 |
| 1 | 52219.52114 | 0.0002 | 6 |
| 2 | 52252.9581 | 0.0003 | 9 |
| 1 | 52263.6290 | 0.00007 | 9 |
| 1 | 52267.8979 | 0.00013 | 9 |
| 1 | 52270.7433 | 0.00014 | 9 |
| 1 | 52287.8168 | 0.0003 | 9 |
| 1 | 52300.6208 | 0.0008 | 8 |
| 2 | 52302.7587 | 0.0004 | 9 |
| 1 | 52317.6964 | 0.0002 | 9 |
| 1 | 52320.5422 | 0.0001 | 8 |
| 2 | 52332.6364 | 0.0002 | 9 |
| 1 | 52337.6167 | 0.0002 | 8 |
| 1 | 52374.6094 | 0.0001 | 8 |
| 1 | 52549.61648 | 0.0003 | 6 |
| 1 | 52563.8448 | 0.00016 | 9 |
| 1 | 52606.5288 | 0.0003 | 10 |
| 1 | 52607.9520 | 0.00018 | 9 |
| 2 | 52615.7779 | 0.0006 | 9 |
| 2 | 52618.6247 | 0.0005 | 8 |
| 1 | 52630.7169 | 0.0003 | 11 |
| 1 | 52683.36123 | 0.0003 | 6 |
| 2 | 52688.34145 | 0.0003 | 6 |
| 2 | 52695.459* | 0.0001 | 12 |
| 1 | 52714.6626 | 0.0002 | 8 |
| 2 | 52729.60329 | 0.00023 | 11 |
| 2 | 52935.91092 | 0.00015 | 11 |
| 1 | 52949.42708 | 0.0002 | 6 |
| 2 | 52971.48018 | 0.0002 | 6 |
| 2 | 52985.7101 | 0.0010 | 13 |
| 1 | 53074.6351 | 0.00011 | 13 |
| 2 | 53079.6147 | 0.00016 | 13 |
| 1 | 53279.51856 | 0.0002 | 6 |
| 2 | 53288.7669 | 0.0006 | 14 |
| 2 | 53315.8018 | 0.00016 | 15 |
| 1 | 53317.93503 | 0.00013 | 15 |
| 1 | 53323.6262 | 0.0006 | 14 |
| 1 | 53354.9284 | 0.0001 | 15 |
| 1 | 53369.1572 | 0.0020 | 16 |
| 1 | 53387.6521 | 0.0016 | 17 |
| 2 | 53388.3641 | 0.0006 | 17 |
| 1 | 53410.4180 | 0.0023 | 17 |
| 1 | 53434.6051 | 0.00011 | 15 |
| 2 | 53452.3913 | 0.0005 | 6 |
| 1 | 53436.0284 | 0.0006 | 18 |
| 1 | 53455.9500 | 0.0020 | 19 |
| 2 | 53462.3529 | 0.0019 | 20 |
| 1 | 53609.6144 | 0.0002 | 21 |
| 2 | 53648.7383 | 0.0008 | 14 |
| 2 | 53674.3522 | 0.0013 | 22 |
| 1 | 53704.9447 | 0.00008 | 23 |
| 1 | 53704.9449 | 0.00010 | 15 |
| 2 | 53709.9231 | 0.0002 | 15 |
| 1 | 53710.6358 | 0.00012 | 15 |
| 1 | 53710.6359 | 0.0001 | 14 |
| 1 | 53713.48108 | 0.0003 | 6 |
| 1 | 53714.9041 | 0.00016 | 15 |
| 2 | 53735.5350 | 0.0003 | 24 |
| 2 | 53758.30032 | 0.0002 | 6 |
| 2 | 53762.5689 | 0.0002 | 14 |
| 1 | 53764.7023 | 0.0003 | 24 |
| 1 | 53771.8161 | 0.0002 | 24 |
| 2 | 53776.7970 | 0.0003 | 24 |
| 2 | 53779.6425 | 0.0002 | 24 |
| 1 | 53781.7770 | 0.0002 | 24 |
| 2 | 53786.7573 | 0.0002 | 24 |
| 1 | 53797.42992 | 0.0003 | 6 |
| 2 | 53802.40763 | 0.0005 | 6 |
| 1 | 53811.6569 | 0.0002 | 24 |
| 2 | 53995.9131 | 0.0003 | 24 |
| 2 | 54015.8334 | 0.0005 | 14 |
| 2 | 54018.6771 | 0.0005 | 25 |
| 2 | 54022.9466 | 0.0003 | 24 |
| 2 | 54032.9059 | 0.0002 | 24 |
| 2 | 54049.9798 | 0.0001 | 24 |
| 2 | 54059.9401 | 0.0001 | 24 |
| 2 | 54069.8998 | 0.0003 | 24 |
| 1 | 54077.7262 | 0.0003 | 24 |
| 2 | 54085.5516 | 0.0023 | 26 |
| 1 | 54091.9536 | 0.0002 | 24 |
| 1 | 54094.7994 | 0.0002 | 24 |
| 1 | 54097.634* | 0.0020 | 14 |
| 2 | 54109.7384 | 0.0004 | 24 |
| 2 | 54115.429 | 0.0008 | 27 |
| 1 | 54117.5646 | 0.0004 | 14 |
| 1 | 54131.7931 | 0.0003 | 24 |
| 1 | 54134.6376 | 0.0002 | 24 |
| 1 | 54134.6378 | 0.0002 | 24 |
| 2 | 54136.7722 | 0.0002 | 24 |
| 2 | 54139.6184 | 0.0001 | 14 |
| 1 | 54161.6709 | 0.0001 | 14 |
| 1 | 54167.3622 | 0.0001 | 28 |
| 2 | 54172.3407 | 0.0005 | 29 |
| 2 | 54172.3419 | 0.0005 | 30 |
| 2 | 54172.3437 | 0.0005 | 31 |
| 1 | 54204.35626 | 0.00040 | 25 |
| 1 | 54204.35696 | 0.00040 | 25 |
| 1 | 54204.35726 | 0.00020 | 25 |
| 1 | 54380.7840 | 0.0002 | 32 |
| 1 | 54387.8986 | 0.0003 | 33 |
| 1 | 54397.8588 | 0.0002 | 33 |
| 1 | 54404.9720 | 0.0002 | 33 |
| 1 | 54404.9722 | 0.0002 | 33 |
| 2 | 54428.4489 | 0.0003 | 31 |
| 2 | 54432.7178 | 0.0002 | 33 |
| 1 | 54447.6580 | 0.0001 | 32 |
| 2 | 54469.7110 | 0.0004 | 33 |
| 2 | 54486.7847 | 0.0007 | 33 |
| 1 | 54487.4952 | 0.0001 | 31 |
| 2 | 54489.6304 | 0.0003 | 33 |
| 1 | 54500.30126 | 0.00010 | 34 |
| 1 | 54500.30136 | 0.00010 | 34 |
| 2 | 54506.7046 | 0.0002 | 32 |
| 1 | 54521.6429 | 0.0002 | 33 |
| 2 | 54522.3559* | 0.0001 | 35 |
| 2 | 54532.31511 | 0.0003 | 6 |
| 2 | 54536.5843 | 0.0004 | 33 |
| 1 | 54737.9104 | 0.0001 | 33 |
| 2 | 54769.9220 | 0.0002 | 33 |
| 1 | 54777.7496 | 0.0002 | 33 |
| 2 | 54779.8821 | 0.0003 | 33 |
| 1 | 54787.7094 | 0.0001 | 33 |
| 2 | 54789.8428 | 0.0003 | 33 |
| 1 | 54834.6609 | 0.0001 | 36 |
| 1 | 54844.6199 | 0.0003 | 37 |
| 2 | 54849.6013 | 0.0002 | 33 |
| 1 | 54857.4262 | 0.0001 | 38 |
| 2 | 54866.6736 | 0.0003 | 33 |
| 1 | 54881.6145 | 0.0002 | 33 |
| 1 | 54888.7294 | 0.0002 | 39 |
| 1 | 54908.6470 | 0.0002 | 33 |
| 2 | 55139.8572 | 0.0003 | 40 |
| 1 | 55147.6829 | 0.0002 | 41 |
| 2 | 55155.5075 | 0.0006 | 35 |
| 1 | 55181.8290 | 0.0002 | 42 |
| 1 | 55184.6746 | 0.0001 | 41 |
| 2 | 55206.7296 | 0.0001 | 41 |
| 1 | 55220.24663 | 0.0003 | 43 |
| 1 | 55241.5885 | 0.0002 | 41 |
| 2 | 55246.5697 | 0.0003 | 41 |
| 1 | 55251.5488 | 0.0007 | 42 |
| 1 | 55258.6627 | 0.0002 | 42 |
| 2 | 55462.8389 | 0.0003 | 42 |
| 2 | 55526.8674 | 0.0001 | 44 |
| 2 | 55596.5852 | 0.0003 | 45 |

| 1 | 55651.36299 | 0.0003 | 6 |
|---|---|---|---|
| 1 | 55803.606 | 0.003 | 46 |
| 1 | 55843.4443 | 0.0003 | 6 |
| 2 | 55855.53767 | 0.0003 | 6 |
| 2 | 55946.5998 | 0.0005 | 47 |
| 1 | 55957.26894 | 0.0003 | 6 |
| 1 | 56220.48903 | 0.0003 | 6 |
| 1 | 56294.47561 | 0.0003 | 6 |
| 2 | 56356.36826 | 0.0003 | 6 |

*These dates were omitted from the final orbital fit due to very bad residuals.

[a]Eclipses of type 1 are the deeper eclipses when the hotter, more massive component (star 1) is being eclipsed by the cooler, less massive component (star 2).

Table 2. CfA radial velocities of HP Aur

| HJD-2,400,000 | $RV_1$ | $RV_2$ | $Err_1$ | $Err_2$ | Phase* |
|---|---|---|---|---|---|
| | | | | | |
| 52532.0148 | 92.60 | -72.30 | 1.59 | 4.26 | 0.6307 |
| 52537.9393 | 119.17 | -102.18 | 1.57 | 4.18 | 0.7946 |
| 52566.9004 | -70.18 | 119.19 | 2.31 | 6.19 | 0.1493 |
| 52569.8456 | -82.32 | 145.77 | 1.75 | 4.69 | 0.2193 |
| 52653.7980 | -83.68 | 134.54 | 1.65 | 4.41 | 0.2235 |
| 52658.8325 | 124.92 | -112.65 | 1.64 | 4.37 | 0.7619 |
| 52685.7124 | 105.02 | -80.02 | 1.50 | 4.02 | 0.6539 |
| 52686.7634 | -47.65 | 96.86 | 1.55 | 4.15 | 0.3926 |
| 52713.7054 | -74.09 | 128.19 | 1.33 | 3.55 | 0.3282 |
| 52718.6213 | 121.45 | -112.72 | 1.93 | 5.15 | 0.7833 |
| 52720.6095 | -77.06 | 124.23 | 1.61 | 4.29 | 0.1806 |
| 52743.6260 | -63.90 | 114.96 | 1.70 | 4.55 | 0.3573 |
| 52745.6289 | 121.85 | -107.36 | 1.62 | 4.33 | 0.7650 |
| 52747.6270 | -71.53 | 123.61 | 1.75 | 4.69 | 0.1693 |
| 52889.9806 | -86.83 | 142.41 | 1.51 | 4.05 | 0.2197 |
| 52958.9859 | 118.23 | -98.66 | 1.67 | 4.46 | 0.7186 |
| 52983.8684 | -81.70 | 137.86 | 1.41 | 3.76 | 0.2068 |
| 52986.8520 | -79.91 | 134.92 | 1.33 | 3.55 | 0.3038 |
| 52988.9342 | 119.00 | -104.45 | 1.68 | 4.50 | 0.7672 |
| 53011.7770 | 113.36 | -93.74 | 1.45 | 3.87 | 0.8218 |
| 53013.8372 | -85.65 | 133.85 | 1.64 | 4.37 | 0.2698 |
| 53016.8065 | -64.67 | 115.17 | 1.81 | 4.85 | 0.3567 |
| 53035.7268 | 104.98 | -87.44 | 1.48 | 3.96 | 0.6544 |
| 53046.6221 | -78.88 | 134.85 | 1.57 | 4.18 | 0.3120 |
| 54072.8470 | 64.93 | -31.26 | 1.27 | 3.39 | 0.5733 |
| 54101.7442 | 91.23 | -71.82 | 1.42 | 3.79 | 0.8831 |
| 54107.8219 | -67.07 | 117.44 | 1.64 | 4.37 | 0.1547 |

*Phase in the inner orbit as computed from the ephemeris in Table 3.

Table 3. Radial velocity/eclipse timing orbital solution

| |
|---|
| ------------------------------------------------------------------------------------------------- |
| INNER ORBIT |
| ------------------------------------------------------------------------------------------------- |
| Adjusted quantities |
| |
| P        1.422819805 ± 0.000000039  days |
| $\gamma$        +19.77 ± 0.15  km/s |
| $K_1$        104.87 ± 0.22  km/s |
| $K_2$        123.66 ± 0.24  km/s |
| Min I    2,453,519.977385 ± 0.000094  (HJD) |
| |
| Derived quantities |
| |
| $M_1 \sin^3 i$        0.9520 ± 0.0041  $M_{sun}$ |
| $M_2 \sin^3 i$        0.8074 ± 0.0036  $M_{sun}$ |
| q            0.8481 ± 0.0025 |

| | | |
|---|---|---|
| $a_1 \sin i$ | $2.0518 \pm 0.0043$ | $10^6$ km |
| $a_2 \sin i$ | $2.4193 \pm 0.0047$ | $10^6$ km |
| $a \sin i$ | $4.4711 \pm 0.0062$ | $10^6$ km |
| $a \sin i$ | $6.4271 \pm 0.0088$ | $R_{sun}$ |
| -------------------------------------------------------------------------------------- | | |
| OUTER ORBIT | | |
| -------------------------------------------------------------------------------------- | | |
| Adjusted quantities | | |
| | | |
| P | $1582.1 \pm 4.0$ | days |
| $K_3$ | $3.46 \pm 0.23$ | km/s |
| e | $0.471 \pm 0.050$ | |
| $\omega$ | $318.3 \pm 4.8$ | deg |
| $T_{peri}$ | $2{,}450{,}403 \pm 20$ | (HJD) |
| | | |
| Derived quantities | | |
| | | |
| $M_3 \sin i$ | $0.1669 \pm 0.0066$ | $(M_1+M_2+M_3)^{2/3}$ $M_{sun}$ |
| $a_{12} \sin i$ | $66.3 \pm 2.6$ | $10^6$ km |
| $a_{12} \sin i/c$ | $0.00256 \pm 0.00010$ | days |
| | | |
| Other quantities pertaining to the fit | | |
| | | |
| $\sigma_{RV}$ (CfA) | $1.5 / 4.0$ | km/s |
| $\sigma_{RV}$ (Popper) | $1.3 / 1.1$ | km/s |
| $N_{RV}$ (CfA/Popper) | $27 / 28$ | |
| $N_{ecl}$ (Min I/Min II) | $101 / 75$ | |
| Time span | $37.2$ | yr |
| -------------------------------------------------------------------------------------- | | |

Table 4.  URSA Differential Photometry of HP Aur

| Orbital Phase | $\Delta V$ | HJD-2400000 |
|---|---|---|
| | | |
| 0.49966 | 1.003 | 52235.88202 |
| 0.50030 | 1.009 | 52235.88294 |
| 0.50093 | 1.021 | 52235.88384 |
| 0.50157 | 1.017 | 52235.88475 |
| 0.50222 | 1.020 | 52235.88567 |

(This table is available in its entirety in a machine-readable form in the online journal.  A portion is shown here for guidance regarding its form and content.)

Table 5. NFO Differential Photometry of HP Aur

| Orbital Phase | $\Delta V^a$ | HJD-2400000 |
|---|---|---|
| | | |
| 0.55061 | 1.633 | 53399.82141 |
| 0.55286 | 1.634 | 53399.82461 |
| 0.55517 | 1.633 | 53399.82790 |
| 0.55741 | 1.632 | 53399.83109 |
| 0.55970 | 1.636 | 53399.83434 |

(This table is available in its entirety in a machine-readable form in the online journal. A portion is shown here for guidance regarding its form and content.)

Note: [a] The NFO differential magnitudes (variable-comparisons) are referenced to the magnitude corresponding to the sum of the flux of two comparison stars.

Table 6. Photometric orbital elements for HP Aur

| Element | URSA | NFO | Adopted |
|---|---|---|---|
| | | | |
| $J_2(V)$ | $0.5479 \pm 0.0098$ | $0.5501 \pm 0.0112$ | $0.549 \pm 0.010$ |
| $r_1 + r_2$ | $0.2809 \pm 0.0005$ | $0.2800 \pm 0.0005$ | $0.2804 \pm 0.0009$ |
| $r_1$ | $0.1599 \pm 0.0003$ | $0.1597 \pm 0.0003$ | $0.1598 \pm 0.0007$ |
| $r_2$ | $0.1210 \pm 0.0003$ | $0.1203 \pm 0.0003$ | $0.1207 \pm 0.0005$ |
| $k$ | $0.7566 \pm 0.0016$ | $0.7530 \pm 0.0018$ | $0.7548 \pm 0.0036$ |
| $i$ (degrees) | $87.716 \pm 0.037$ | $87.679 \pm 0.042$ | $87.698 \pm 0.040$ |
| $e$ | 0 fixed | 0 fixed | 0 |
| $u_1$ | $0.111 \pm 0.025$ | $0.193 \pm 0.027$ | $0.15 \pm 0.04$ |
| $u_2$ | $0.043 \pm 0.052$ | $0.142 \pm 0.059$ | $0.09 \pm 0.05$ |
| $y_1$ | 0.038 fixed | 0.038 fixed | 0.038 |
| $y_2$ | 0.039 fixed | 0.039 fixed | 0.039 |
| q | 0.8481 fixed | 0.8481 fixed | 0.8481 |
| $L_3$ | 0 fixed | 0 fixed | 0 |
| $L_1$ | 0.7575 ± 0.0029 | 0.7595 ± 0.0028 | 0.7585 ± 0.0021 |
| $L_2$ | 0.2425 ± 0.0029 | 0.2405 ± 0.0028 | 0.2415 ± 0.0021 |
| $L_2/L_1$ (V) | 0.3201 ± 0.0011 | 0.3166 ± 0.0014 | 0.318 ± 0.004 |
| $\sigma$ (mmag) | 11.26821 | 6.91833 | |
| N | 6685 | 2472 | |
| Corrections | 132 | 73 | |

Table 7.  Absolute properties of the close binary stars in HPAur

| Parameter | Star 1 | Star 2 |
|---|---|---|
| Mass (solar masses) | 0.9543 ± 0.0041 | 0.8094 ± 0.0036 |
| Radius (solar radii) | 1.0278 ± 0.0042 | 0.7758 ± 0.0034 |
| log g (cm s$^{-2}$) | 4.3942 ± 0.0040 | 4.5669 ± 0.0043 |
| Eccentricity e | 0* | |
| v sin i (km s$^{-1}$) (observed value) | 41 ± 3 | 30 ± 5 |
| Circular $v_{sync}$ (km s$^{-1}$) (equatorial) | 36.5 ± 0.1 | 27.6 ± 0.1 |
| Orbital semi-major axis a | 6.432 ± 0.009 | |

| (solar radii) | | |
|---|---|---|
| $T_{eff}$ (K) | 5810 ± 120 | 5160 ± 120 |
| log L (solar units) | 0.036 ± 0.035 | -0.414 ± 0.039 |
| $M_V$ (mag) | 4.68 ± 0.12 | 5.94 ± 0.10 |
| $F_V$ | 3.752 ± 0.012 | 3.687 ± 0.010 |
| $E_{B-V}$ reddening (mag) | 0.048 ± 0.020 | |
| m-M (mag) | 6.61 ± 0.15 | |
| Distance (pc): | | |
| from Popper (1980) | 214 ± 13 | |
| from Flower (1996) | 209 ± 14 | |

*The eccentricity e, when allowed to be a variable, was found to be 0.0006 ± 0.0016, so it was fixed at 0 in the final orbital fit.